# Observability of modified threshold behavior near unitarity

Michael D. Higgins[1], J. Golak[2], R. Skibiński[2], K. Topolnicki[2], H. Witała[2], H. Kamada[3,4], Chris H. Greene[1,5]

[1*]Department of Physics and Astronomy, Purdue University, West Lafayette, Indiana 47907 USA.
[2]M. Smoluchowski Institute of Physics, Faculty of Physics, Astronomy and Applied Computer Science, Jagiellonian University, PL-30348 Kraków, Poland.
[3]Department of Physics, Faculty of Engineering, Kyushu Institute of Technology, Kitakyushu 804-8550, Japan.
[4]Research Center for Nuclear Physics, Osaka University, 10-1 Mihogaoka, Ibaraki, Osaka 567-0047, Japan.
[5]Purdue Quantum Science and Engineering Institute, Purdue University, West Lafayette, Indiana 47907 USA.

Contributing authors: higgin45@purdue.edu ; jacek.golak@uj.edu.pl ; roman.skibinski@uj.edu.pl ; kacper.topolnicki@uj.edu.pl ; henryk.witala@uj.edu.pl ; kamada@rcnp.osaka-u.ac.jp, kamada@mns.kyutech.ac.jp ; chgreene@purdue.edu ;

**Abstract**

A number of recent references have pointed out that an $N$-particle system having short-range interactions at $S$-wave and/or $P$-wave unitarity can exhibit modified threshold behavior for various reactive processes. But the question of how close to unitarity one must get in order to observe such modifications has not been addressed. The present study quantifies this question by treating cases involving 3- or 4-neutrons, at the physical value of the neutron-neutron singlet scattering length $a$ and at artificially altered values. One major conclusion is that the neutron-neutron scattering length is not yet sufficiently large for the $3n$ or $4n$ systems to demonstrate the unitarity threshold exponent.

# 1 Introduction

One development that has been increasingly discussed in the few-body physics community in recent years is the fact that when a two-body $S$-wave scattering length and/or $P$-wave scattering volume diverge, the relevant Wigner-type threshold law [1] for various processes can be altered from its behavior for finite two-body interactions. An example of this modification occurs when the initial or final state lies in the 3-body scattering continuum and the initial or final state energy, respectively, lies very close to the threshold energy for the 3 particles to arrive from or to escape to infinity. One reference that has discussed this threshold law modification for processes involving 3 or 4 fermions is an article by Hammer and Son [2], who derive their result using arguments from conformal field theory. Another set of recent references [3–5] has documented many different threshold law modifications at $S$-wave and/or $P$-wave unitarity, for a number of different symmetries of 3-fermion and 4-fermion systems. Those latter references reached these conclusions by basing their argument on the relevant channel potential energy curves that were computed in the adiabatic hyperspherical representation. Moreover, they go on to argue that the threshold law modifications can be viewed as a generalized manifestation of Efimov physics, using arguments that are summarized below.

The original argument in the aforementioned 1948 article by Eugene Wigner was based on an analysis of the effective long-range potential in any near-threshold fragmentation channel; specifically, when the long-range (rescaled) radial Hamiltonian in the $i$-th such channel has the following asymptotic form,

$$T_r + V_i(r) \to -\frac{\hbar^2}{2\mu}\frac{\partial^2}{\partial r^2} + \frac{\ell_i(\ell_i+1)\hbar^2}{2\mu r^2}, \tag{1}$$

this results in the near-threshold energy dependence of a squared transition amplitude in such a channel that has the form

$$|T_{ii'}(E)|^2 \propto k_i^{2\ell_i+1}, \tag{2}$$

where $k_i \equiv [2\mu(E-E_i)/\hbar^2]^{1/2}$ is the wavenumber in the near-threshold channel $i$. Note that $\mu$ is the effective mass that appears in the kinetic energy operator, assumed here to act on the rescaled radial wavefunction as a pure second derivative term with no diagonal first derivative terms present. This discussion of the threshold law is based on the assumption that the near-threshold channel is the initial state channel, as in the meson capture processes that are the focus of the examples below. But for an $N$-body recombination reaction, where it is the final state channel that is the one near threshold, one replaces $k_i$ and $\ell_i$ by $k_{i'}$ and $\ell_{i'}$. Moreover, in the special case that $k_i$ and $k_{i'}$ are degenerate, both the initial and final Wigner threshold factors will be present in $|T_{ii'}(E)|^2$. One other point to keep in mind is that while $\ell_i(\ell_i+1)$ is a convenient way to write the coefficient of the asymptotic $r^{-2}$ potential, it does not necessarily represent an eigenvalue of a squared angular momentum operator; in other words, it can in some cases incorporate particle interactions as well, as in the Efimov effect in particular and in the unitarity limits discussed below. For this reason, we will not refer to $\ell_i$ as an angular momentum quantum number, but instead will denote it as an "effective centrifugal constant".

In the framework of the hyperspherical representation utilized here, the key quantity $\ell_{\text{eff}}$ controlling the Wigner threshold law emerges in the same manner as in Wigner's original argument, namely by finding the coefficient of $r^{-2}$ in the near-threshold channel of interest. Note that one earlier article [6] also computed the values of $\ell_{\text{eff}}$ used below in our four-fermion threshold law analysis, although that reference did not discuss the implication that there would be modified threshold laws of reactive processes at unitarity. In certain circumstances, however, the value of $\ell_{\text{eff}}$ can be deduced indirectly through a bound state energy level calculation. This idea has been implemented in studies of two-component fermionic systems at unitarity by Werner and Castin [7] and similarly by Yin and Blume [8]. The idea is to add a symmetric hyperradial harmonic oscillator potential to the Hamiltonian, i.e. $\frac{1}{2}\mu\omega^2 r^2$. Then, *provided the centrifugal $r^{-2}$ term is known to be repulsive or can be assumed to be*, the full potential energy is dominated by the centrifugal and oscillator terms, and the energy levels after removing the center-of-mass energy can be expressed analytically in terms of $\ell_{\text{eff}}$ and the number $n_r$ of radial nodes, namely $E(n_r, \ell_{\text{eff}}) = (2n_r + \ell_{\text{eff}} + 3/2)\hbar\omega$. After numerically computing the energy levels, e.g. variationally or using Monte Carlo methods, the oscillator level formula can be inverted to yield the value of $\ell_{\text{eff}}$.

We will not repeat the arguments and calculations of Refs. [3–6], nor of Refs. [7, 8], which have determined the lowest values of $\ell_{\text{eff}}$ in multiple symmetries of various systems, but the reader is encouraged to explore those references. Instead, Table 1 reproduces several of the $\ell_{\text{eff}}$ values that will be needed in our discussion below about the near-threshold energy dependence of various processes. In the following, keep in mind that for any point away from unitarity, for any squared transition amplitude with either a near-threshold bra or ket, the generalized Wigner law energy dependence sufficiently close to threshold must be $|T_{f,i}|^2 \propto (E - E_{\text{th}})^{\ell_{\text{eff}}^{\text{NI}} + \frac{1}{2}}$, where $\ell_{\text{eff}}^{\text{NI}}$ is the lowest value of that effective orbital angular momentum in the near-threshold channel. On the other hand, if one tunes the two-body scattering length exactly to unitarity, $a \to \infty$, one would observe a different threshold law governed by the value of $\ell_{\text{eff}}^{\text{U}}$ from Table 1. In order to connect with conformal field theory, observe that measurable quantities such as the differential cross section scale near threshold as $d\sigma/dE \approx (E - E_{\text{th}})^{\Delta - 5/2}$, where $\Delta$ is the conformal dimension. [2] Moreover, $\ell_{\text{eff}}$ is related to $\Delta$ through the equality $\Delta = \ell_{\text{eff}} + 3$.

A major conclusion documented quantitatively in the following sections of the present study, however, is that *observation of the unitarity-limit of the threshold law is possible only when one is extremely close to unitarity*. Moreover, our analysis leads to a different conclusion than that of Hammer and Son [2], who analyzed two theoretical calculations of 3-neutron formation by Golak and collaborators [9–11]: the evidence presented below shows that the physical value of the singlet neutron-neutron scattering length (-18.5 fm) is not sufficiently close to unitarity to realistically claim that the unitarity-modified threshold law is observable in any short-range reactive process leading to 3-neutron or 4-neutron formation. This conclusion is supported by calculations of trineutron formation in the reaction of a triton and a negative pion carried out by Golak *et al.* [9] plus additional new previously unpublished calculations carried out by those authors at lower energies. Similarly, we present a hyperspherical calculation that suggests a similar conclusion for any short-range reactive processes that could form the 4-neutron system at low energies. Only if the neutron-neutron scattering length were approximately 3 to 5 times larger than its currently known value of -18.5 fm would the unitarity-modified threshold law be realistically observable for the 3n or 4n system.

Of even greater interest, however, than the question of whether the unitarity-modified threshold law "has" or "has not" been observed for a process, the present article discusses the way that the ordinary non-unitary threshold law will always be valid at sufficiently low energies, since the scattering length will never be "exactly" infinite, but there will be a transition region between the ordinary threshold law and the unitarity-modified law when the two-body interactions are sufficiently close to unitarity, and we document that transition in this report.

Table 1 Values of the effective angular momentum quantum number for the non–interacting case $\ell_{\text{eff}}^{\text{NI}}$ and at the $s$–wave unitarity limit $\ell_{\text{eff}}^{\text{U}}$ are given for the ground– and first–excited states of $N = 3$ and $N = 4$ fermionic systems. The value of the symmetry dependent $a\rho^{-3}$ coefficient $C$ in Eq. (3) is also given.

| N | (Spin State) $L^\pi$ | $\ell_{\text{eff}}^{\text{NI}}$ | $\ell_{\text{eff}}^{\text{U}}$ | $C$ |
|---|---|---|---|---|
| 3 | $(\uparrow\uparrow\downarrow)1^-$ | 5/2 | $1.2727^1$ | $15.1(3)^2$ |
| 3 | $(\uparrow\uparrow\downarrow)0^+$ | 7/2 | $1.668^1$ | $29.8(5)^2$ |
| 4 | $(\uparrow\uparrow\downarrow\downarrow)0^+$ | 5 | $2.017^1$ | $86.7(3)^2$ |
| 4 | $(\uparrow\uparrow\downarrow\downarrow)1^-$ | 6 | $4.11(2)^1$ | $73.6(3)^2$ |

[1]Values extracted from Table 3 of [4] and Table 2 of [12].

[2]Values extracted from Table 3 of [12].

## 2 Theoretical Methods

For a system of $N$ fermions with short-range $S$-wave interactions characterized by a scattering length $a$, a number of studies have demonstrated that the leading terms in the lowest adiabatic hyperradial potential curve converging to the $N$-body continuum threshold energy (taken to be the zero of our energy scale) has the following form at large hyperradii $\rho$:

$$W_0(\rho) \xrightarrow[\rho \to \infty]{} \frac{\ell_{\text{eff}}^{\text{NI}}(\ell_{\text{eff}}^{\text{NI}} + 1)\hbar^2}{2\mu\rho^2} + C\frac{a\hbar^2}{2\mu\rho^3}, \qquad (3)$$

where $C$ is a dimensionless constant that depends on the number of particles, their identical particle statistics, and on their symmetry quantum numbers, and $\mu$ is the nucleon–nucleon reduced mass.

In the limit where the two-body scattering length $a$ is "exactly" infinity, the asymptotic potential energy in Eq.3 is no longer valid, and a new form applies with an "Efimov-physics-modified" unitarity value for the centrifugal term, and with no $\rho^{-3}$ term at all,

$$W_0(\rho) \xrightarrow[\rho \to \infty]{} \frac{\ell^{\text{U}}_{\text{eff}}(\ell^{\text{U}}_{\text{eff}} + 1)\hbar^2}{2\mu\rho^2} + O(\rho^{-4}). \tag{4}$$

Note that Eqs. (3) and (4) also hold for the problem of 3 identical bosons, where the most attractive channel in the 3-body continuum (with $L^\pi = 0^+$) has $\ell^{\text{NI}}_{\text{eff}} = \frac{3}{2}$ and $\ell^{\text{U}}_{\text{eff}} = -\frac{1}{2} + is_0$. The universal constant $s_0 \approx 1.00624$ leads to a negative long range $\rho^{-2}$ potential at unitarity, and a qualitative difference between the spectrum at finite $a$ where there are at most a finite number of bound or resonant states, and $a \to \infty$ where there are an infinity of bound states. This is the prototypical case of the Efimov effect. Other systems that show signatures of Efimov physics at unitarity, but without the infinity of levels that represent the Efimov *effect*, are the 3n and 4n systems which are the focus of the present analysis.

In particular, at exact unitarity, where $a \to \infty$, solutions of the $N$-body Schrödinger equation with two-component fermions have been carried out in various references for $N = 3$ and $N = 4$ for different symmetries, in order to determine the effective centrifugal constant $\ell^{\text{U}}_{\text{eff}}$. Table 1 shows several values of the reduced $\ell^{\text{U}}_{\text{eff}}$, which determine the modified Wigner threshold law exponents in that limit for those systems.

One learns two things about near-threshold behavior in such a long-range channel potential, e.g. from the masterful review article [13]:

*(i)* The near-threshold elastic scattering phaseshift in such a potential with this behavior at long range has the following leading-order behavior as $k \to 0$:

$$\delta_{\ell^{\text{NI}}_{\text{eff}}}(k) \to -\frac{Cak}{2\ell^{\text{NI}}_{\text{eff}}(\ell^{\text{NI}}_{\text{eff}} + 1)}. \tag{5}$$

This low energy phaseshift behavior (valid provided $\ell^{\text{NI}}_{\text{eff}}$ is real, nonzero, and positive) was discussed in Refs. [12, 14] and is interesting because it produces a divergent Wigner-Eisenbud-Smith time delay that can also be viewed as a density of states enhancement (nonresonant) near threshold in the 3- and 4-neutron systems. However, this phaseshift behavior *(i)* derives from long-range behavior [13], and thus is not relevant to the present study, which instead involves the second, short-range aspect of threshold law physics, namely:

*(ii)* The leading near-threshold behavior of a squared transition amplitude governed by any short-range operator will behave near threshold as:

$$|T_{f,i}|^2 \xrightarrow[k \to 0]{} \text{constant} \times k^{2\ell^{\text{NI}}_{\text{eff}} + 1}, \tag{6}$$

*provided* $|a| < \infty$. As was mentioned above, $k$ in this formula is understood to be either the initial ($k_i$) or the final ($k_f$), depending on whether the initial or the final wavenumber is the one near threshold. The energy range of validity of this threshold behavior is expected to hold at least up to the range $k \ll (Ca)^{-1}$, and in practice well beyond that point. For a few-body system at exact unitarity, Eq. (6) holds except with the replacement $\ell_{\text{eff}}^{\text{NI}} \to \ell_{\text{eff}}^{\text{U}}$.

Another point worth making about Wigner-type threshold laws is that they are, strictly speaking, only giving the dominant power-law behavior in the limit of the energy going to threshold, which we write throughout this article as the limit $E \to 0$. That threshold law energy dependence is understood to multiply an energy dependent function that can be written as an analytic function of energy with a nonvanishing radius of convergence, i.e., $F(E) = c_0 + c_1 E + c_2 E^2 + ...$, which approaches a constant in the $E \to 0$ limit. Thus an expression of the near-threshold behavior that has validity over a broader energy range than Eq. (6) is the following:

$$|T_{f,i}|^2 \xrightarrow[k \to 0]{} c_0 \ k^{2\ell_{\text{eff}}^{\text{NI}}+1}(1 + c_1 E + c_2 E^2 + ...) \tag{7}$$

The Jeffries-Wentzel-Kramers-Brillouin (JWKB) approximation [15] is in fact sufficiently accurate to give an intuitive understanding of the fact that a simple tunneling integral controls most of the physics of the generalized Wigner-type threshold behavior of interest here. A first order JWKB treatment, in particular, describes the rescaled hyperradial wave function $\psi \equiv \rho^{\frac{3N-4}{2}}\Psi$, normalized to have an asymptotic amplitude that is energy-normalized; it has the following form in the short-range classically forbidden region:

$$\psi_{\text{JWKB}}(k, \rho_{\text{small}}) = \sqrt{\frac{\mu}{2\pi\hbar|p(k,\rho_{\text{small}})|}}\exp\left[-\frac{1}{\hbar}\int_{\rho_{\text{small}}}^{\rho_{\text{max}}} d\rho' \ \text{Im}[p(k,\rho')]\right]. \tag{8}$$

This equation is valid for a potential energy curve that is monotonically decreasing and has only one classical turning point at all positive energies, which was shown to apply to the $3n$ and $4n$ systems in Refs. [12, 14]. Here $k$ is the wave number defined in terms of the energy $E$ as $k = \sqrt{2\mu E/\hbar^2}$, $\rho_{\text{small}}$ is the hyperradius inside the classically forbidden region where the wave function is evaluated, $\rho_{\text{max}} \approx (\ell_{\text{eff}} + \frac{1}{2})/k$ is the location of the outermost classical turning point. The local Langer-corrected momentum $p(k, \rho)$ is defined as:

$$p(k, \rho) = \sqrt{2\mu\left(E - W_0(\rho) - \frac{\hbar^2}{8\mu\rho^2}\right)}. \tag{9}$$

In Eq. (9), $p(k, \rho)$ is defined as an effective local momentum that depends on the hyperradius through the hyperradial potential energy curve $W_0(\rho)$ and which includes the Langer correction term $\hbar^2/8\mu\rho^2$ for purely radial potentials. At low energies, the wave function given in Eq. (8) has a power–law dependence on the wave number $k$ given as $\Psi_{\text{JWKB}}(k, \rho_{\text{small}}) \propto k^{\ell_{\text{eff}}+1/2}$. The squared JWKB tunneling amplitude for tunneling in the classically forbidden region is given as:

$$\tau_{\text{JWKB}}(k, \rho_{\text{small}}) = \exp\left[-\frac{2}{\hbar}\int_{\rho_{\text{small}}}^{\rho_{\text{max}}} d\rho' \ \text{Im}[p(k,\rho')]\right]. \tag{10}$$

The squared tunneling amplitude in Eq. (10) has the same threshold law dependence on the wave number at low energies, specifically $\tau_{\text{JWKB}}(k, \rho_{\text{small}}) \propto k^{2\ell_{\text{eff}}+1}$. To map out the power law behavior of the wave function and tunneling amplitudes, it is convenient to write the near-threshold energy dependence in terms of an energy-dependent effective exponent $\gamma(k)$, which is defined at finite energy through an assumed proportionality of $\psi(k, \rho_{\text{small}}) \propto k^{\gamma(k)}$ at small $k$. Thus, to assess the energy dependence of the exponent $\gamma(k)$ in a numerically computed short-range wavefunction, the functional form of $\gamma(k, \rho_{\text{small}})$ becomes

$$\gamma(k, \rho_{\text{small}}) = k \frac{d}{dk} ln|\Psi(k, \rho_{\text{small}})|. \tag{11}$$

Of course this effective exponent formula for $\gamma(k, \rho_{\text{small}})$ only represents an approximate power law $k$-dependence if the $k$-dependence of $\gamma(k)$ is very weak. But with this assumption, which will be seen to hold in the test cases discussed below, comparisons can now be made to the expected limiting value, $\gamma = \ell_{\text{eff}} + \frac{1}{2}$, which must hold at threshold for any non-infinite value of the scattering length $a$.

It is instructive to view the low-energy, near-threshold behavior of the transition matrix element of a short-range operator as having its energy dependence controlled by three distinct regions: Region *(I)* In the lowest energy region closest to threshold, the *shape* of the final state wavefunction in the region of the matrix element $\langle E_f|\hat{T}|E_i\rangle$ is energy-independent to an excellent approximation, and essentially all of the energy dependence derives from the squared WKBJ tunneling amplitude in Eq. (10). Moreover, in this energy range the outermost turning point $\rho_{\text{max}}$ is sufficiently large that the approximation $\rho_{\text{max}} \approx (\ell_{\text{eff}}^{\text{NI}} + \frac{1}{2})/k$ holds quite accurately and therefore $\tau_{\text{JWKB}} \propto k^{\ell_{\text{eff}}^{\text{NI}} + \frac{1}{2}}$ also holds accurately. In the next range of energies, Region *(II)*, the shape of the final state wavefunction in the small-$\rho$ region of the transition operator matrix element is still approximately energy-independent, but now the outermost turning point $\rho_{\text{max}}$ is at smaller distances and it lies in the region where the correction terms to the potential are non-negligible. In that energy range *(II)*, the transition matrix element energy dependence begins to deviate from $k^{\ell_{\text{eff}}^{\text{NI}} + \frac{1}{2}}$, but the new energy-dependence is still controlled by the squared JWKB tunneling amplitude in Eq. (10). Finally, in the uppermost range of near-threshold energies, Region *(III)*, the actual shape or nodal structure of the final state wavefunction in the small-$\rho$ region of the transition operator matrix element is beginning to change with energy, and as a result the transition matrix element begins to vary in a possibly complicated manner that usually cannot be described as a modified power law. The behavior of transition rates in these three different regions is documented in the analysis below.

To compare results with the approximate JWKB wave function, accurate numerical solutions to the hyperradial Schrödinger equation have also been computed. The solution to the $N$–body Schrödinger equation in hyperspherical coordinates in the adiabatic approximation for the lowest channel $\nu = 0$ is given as $\Psi(\rho, \Omega) \approx \rho^{-\frac{3N-4}{2}} F_0(\rho) \Phi_0(\rho, \Omega)$, where $\Phi_0(\rho, \Omega)$ is the lowest hyperspherical channel function in the $N$-body continuum, obtained by diagonalizing the adiabatic (fixed-$\rho$) Hamiltonian at fixed hyperradius [12, 14]. The hyperradial function $F_0(\rho)$ is found by solving the hyperradial Schrödinger equation,

$$-\frac{\hbar^2}{2\mu}\frac{d^2}{d\rho^2}F_0(\rho) + W_0(\rho)F_0(\rho) = EF_0(\rho) \qquad (12)$$

where $E$ is the energy of the system, $W_0(\rho)$ is the lowest adiabatic hyperradial potential energy curve, and $\mu$ is the hyperradial reduced mass of the system.

## 3 Results

As Hammer and Son realized, the high quality solutions of the Fadeev equations for the processes $^3$H$(\pi^-,\gamma)3n$, $^3$H$(\mu^-,\nu_\mu)3n$, and $^3$H$(\bar{\nu},e^+)3n$ by Golak and collaborators [9–11] provide an opportunity to explore the near-threshold behavior of a system close to unitarity, namely the system of 3 neutrons. Their calculated energy distribution of the final states produced in the pion capture process was shown to agree with experimental measurements, and this adds to our confidence that their calculation can be used to realistically explore the near-threshold energy dependences implied by those numerical results.

In the hyperspherical treatment of the $3n$ or $4n$ final state, our focus is on the energy-dependence of the wavefunction at small hyperradii, in the near threshold energy range. The primary focus of our analysis is in the lower two out of the three energy regions described above, in particular those where the main energy dependence is in the overall scale factor of the wavefunction, as opposed to an energy-dependent change in the final state nodal structure. One criterion for a system where the scattering length is the dominant length scale in the underlying two-body interactions, and that criteria states that the threshold law $E^{\ell_{\text{eff}}^{\text{NI}}+\frac{1}{2}}$ based on the noninteracting value $\ell_{\text{eff}}^{\text{NI}}$ of the effective centrifugal constant applies for $k \lesssim 1/|a|$. When $|a|$ is large, this will be an extremely small energy range, of course. The second criterion for the relevant range where one could observe a modified threshold law is over the range $k \lesssim 1/r_0$ where $r_0$ is the range of the two-body potential energy, i.e. $r_0 \approx 2.73$ fm for the neutron-neutron system. This second range where one can observe near-threshold behavior that is potentially modified when approaching the unitarity limit $|a| \to \infty$.

Two different choices for the two–body neutron–neutron interactions have been used in the present study, namely, the Argonne AV8$'$ interaction [16] and a single Gaussian two-body interaction for the spin singlet and triplet states, which has been shown [12, 14] to produce results very close to realistic nucleon–nucleon Argonne potentials. The single Gaussian interactions are used when tuning the $s$–wave interaction to give a scattering length that differs from the bare neutron–neutron scattering length. The adiabatic hyperradial potential curves are computed by diagonalizing the adiabatic Hamiltonian at fixed hyperradius using an explicitly correlated Gaussian (CGHS) basis [17].

### 3.1 3n System

Figure 1 shows the lowest adiabatic hyperspherical potential curve $W_0(\rho)$ for the single Gaussian representation of the neutron–neutron two–body interaction in the $J^\pi = 3/2^-$ symmetry. The different solid curves corresponds to different neutron–neutron $s$–wave scattering lengths ranging from $1\times$ to $50\times$ the neutron–neutron scattering length, which is $a_{nn} = -18.5$ fm [16]. The scattering length is tuned by varying the overall strength of the spin–singlet interaction, while the spin–triplet interaction remains unchanged. For reference, the non-interacting potential and the unitarity limiting potential of the form $W(\rho) = \frac{\hbar^2 \ell_{\text{eff}}(\ell_{\text{eff}}+1)}{2\mu\rho^2}$ are also shown. In the $J^\pi = 3/2^-$ symmetry, $\ell_{\text{eff}}^{\text{NI}} = 5/2$ for the non–interacting (or finite $a$) case, while $\ell_{\text{eff}}^{\text{U}} = 1.2727$ for the unitarity limiting case (see Table 1). These potentials are shown by the dashed and dot–dashed lines, respectively. As the spin–singlet scattering length is increased to more negative values, i.e. multiples of the bare neutron–neutron scattering length, the repulsive hyperradial potential becomes more attractive and gets closer to the unitarity limiting potential that applies at $a_s \to -\infty$. The reduction in $\ell_{\text{eff}}$ as the scattering length gets more and more negative towards the unitarity limit has implications for the threshold law behavior of the relevant off-diagonal elements of the scattering matrix $\hat{S}$ or the transition matrix $\hat{T}$, which vary in proportion to $k^{\ell_{\text{eff}}+1/2}$ at small wave numbers. In practice, the adiabatic hyperspherical potential curves are fitted to their asymptotic form at intermediate hyperradii, and used at large hyperradii. The Fig.1 potential curve fits give the following form for $\rho > 10$ fm,

$$W_0^{(3n)}(\rho) = \frac{\hbar^2}{2\mu\rho^2}\left(\ell_{\text{eff}}^{\text{NI}}(\ell_{\text{eff}}^{\text{NI}}+1) + \frac{Ca_{nn}}{\rho} + \frac{C_2}{\rho^2} + \frac{C_3}{\rho^3} + \frac{C_4}{\rho^4} + \frac{C_5}{\rho^5} + \frac{C_6}{\rho^6}\right). \quad (13)$$

For the case where the scattering length is the bare neutron–neutron $^1S_0$ value denoted $a_{nn}$, the expansion coefficients for the $3n$ hyperradial potential in this most attractive symmetry are $C = 15.1$, $C_2 = 9,532(5)$ fm$^2$, $C_3 = -201.0(2) \times 10^3$ fm$^3$, $C_4 = 2.515(4) \times 10^6$ fm$^4$, $C_5 = -1.688(4) \times 10^7$ fm$^5$, and $C_6 = 4.66(2) \times 10^7$ fm$^6$.

Figure 2 shows the three neutron relative energy spectrum from [10], represented as the differential capture rate $d\Gamma/dE$, for the pion capture process $^3\text{H}(\pi^-,\gamma)3n$ as a function of the three–neutron energy. Their predictions are represented by the open circles and is compared to the squared of the JWKB wavefunction (dot–dashed curve), rescaled by an overall constant factor to obtain the same order of magnitude of their data. The JWKB wave function is evaluated at $\rho_{\text{small}} = 1.25$ fm, which is well inside the classically forbidden region. It is relevant that the hyperspherical potential curve for the triton, shown in Fig. 1 of Ref. [18], implies that the initial state wavefunction will only extend out to around a hyperradius $\rho \approx 5$ fm. We have confirmed that the energy dependence is identical, aside from an overall constant scaling factor, if we use different values of $\rho_{\text{small}}$ in the range from 1.25 fm to 5 fm. The purpose of comparing the data for $d\Gamma/dE$ with the JWKB wave function is to highlight the various threshold law behaviors at low energies. At low energies, the wave function and the quantity $d\Gamma/dE$ scale as $E^3$, which is consistent with the ordinary Wigner threshold law of $k^{2\ell_{\text{eff}}+1}$ with $\ell_{\text{eff}} = 5/2$. This is exactly what is expected from an adiabatic hyperspherical potential that approaches, at large hyperradius, $W_0 \to \frac{\hbar^2 \ell_{\text{eff}}(\ell_{\text{eff}}+1)}{2\mu\rho^2}$ with the same value of $\ell_{\text{eff}} = 5/2$ that applies in the non–interacting limit and which also represents the dominant long-range term asymptotically at $|a| < \infty$. Fig.2 shows that the energy-dependent scale factor from the JWKB wave function in the hyperradial potential agrees with the data for $d\Gamma/dE$ up to about $E \approx 2$ MeV. At higher energies, in the range 5 MeV $< E <$ 10 MeV, the JWKB scale factor deviates from the capture rate energy depencence. Above about 2 MeV the energy dependence is no longer controlled merely by an overall energy-dependent scale factor in the short-range wavefunction; as was argued above, this higher energy range is Region *(III)* where the nodal structure begins to affect the energy-dependence of $d\Gamma/dE$, as we will document below (see the discussion below, relating to Fig. 4).

It could be tempting to argue (as in Ref. [2]) that the energy dependence in the region between approximately 1 and 3 MeV starts to have a slope consistent with the unitarity limiting exponent, $\ell^U_{\text{eff}} + \frac{1}{2} = 1.7727$, but we view that apparent agreement over a limited range as somewhat fortuitous. One way to view the effective energy exponent as it evolves with energy is shown in Fig. 3, which displays the exponent computed using Eq. (11). That graph only goes up to 2 MeV, since the energies above that are affected by the integrand containing the initial state, the energy-dependent final state, and the transition operator (corresponding to Region *(III)* in the discussion below Eq. (11). The effective exponent only gets within 20% of the unitarity value at 2 MeV.

Fig. 3 displays the effective threshold exponent $\gamma(k, \rho_{small})$, plotted versus the energy using the JWKB wave function evaluated at a hyperradius of $\rho_{\text{small}} = 1.25$ fm. Each curve represents a different neutron–neutron spin–singlet scattering length, tuned to 1, 5, 10, and 50 times the bare neutron–neutron scattering length. The two horizontal lines mark the bounds on the exponent, with one at $\gamma = 3.0$, marking the non–interacting value, and the other at $\gamma = 1.7727$, marking the unitarity limiting value. As the scattering length increases, the effective threshold exponent gets closer to 1.7727, over a broader energy range, as expected since the *s*–wave interaction approaches unitarity; that is the point where the two-body singlet scattering length becomes infinite or equivalently, where the neutron-neutron system acquires a bound state with vanishing binding energy. Again, as was pointed out above, the $\gamma$ computed using Eq. (11) can only truly be interpreted as an "effective threshold exponent" where the energy dependence is not too strong, like at very low energy or at higher energies. The small deviations from the unitarity limit at higher energies are not particularly relevant to our near-threshold analysis that is the focus of the present study. Our conclusion from inspection of Fig. 3 is that the singlet neutron-neutron scattering length would have to be at least 3-5 times more negative in order to observe the unitarity-modified threshold law over an appreciable energy range.

The higher energy range in Region (*III*) is studied more carefully by analyzing the energy dependence of the overlap of the 3*n* wavefunction with the triton wavefunction for different values of the triton binding energy. The triton wavefunction is computed using the Minnesota nucleon–nucleon interaction with a purely hyperradial three–body force, which is included to modify the binding energy without changing the two–body physics. The quantity $|\langle \psi_{^3\text{H}} | \psi_{3n} \rangle|^2$ is computed as a function of energy for two different artificial values of the triton binding energy and compared with the capture rate $d\Gamma/dE$ for the process $^3\text{H}(\pi^-, \gamma)3n$, and shown in Fig. 4. As the binding energy decreases the energy dependence in $|\langle \psi_{^3\text{H}} | \psi_{3n} \rangle|^2$ in Region (*III*) begins to have a larger curvature and turns over, capturing the energy dependence seen in $d\Gamma/dE$ at

energies greater than 5 MeV. This indicates that it is the nodal structure of the triton wavefunction that controls the energy dependence at large energies. At small energies in Region (I), the low–energy threshold law behavior kicks in and is unaffected by the triton binding energy and scales as $E^3$ for energies less than 1 MeV as indicated by the dashed line in Fig. 4.

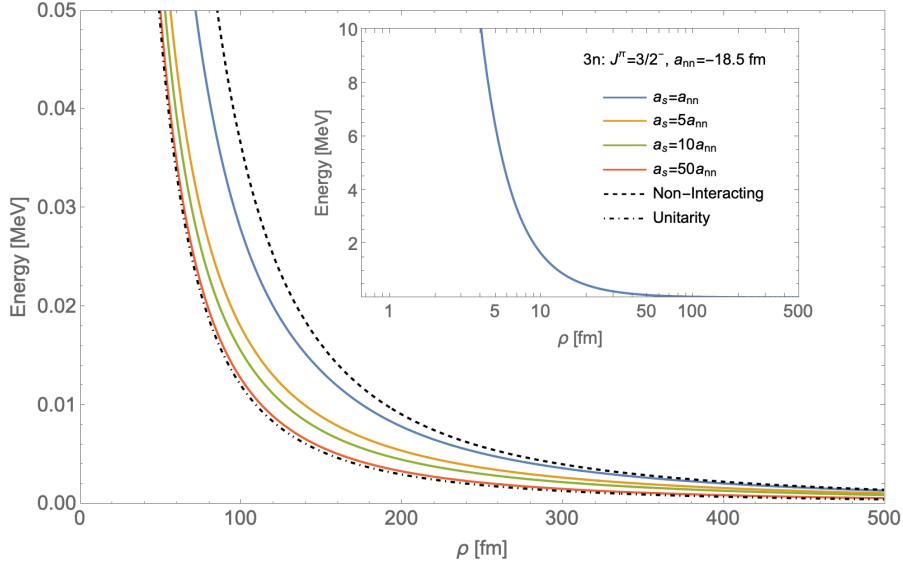

**Fig. 1** This plot shows the lowest hyperradial potential curve for the 3n system in the $J^\pi = 3/2^-$ symmetry using a Gaussian representation for the spin–singlet and triplet two–body interactions [12, 14]. The neutron–neutron scattering length is tuned to different values by an overall rescaling of the spin–singlet interaction. The value of the neutron–neutron scattering length is $a_{nn} = -18.5$ fm. The potential curves for different scattering lengths are represented by the solid lines and corresponds to scattering lengths (from top to bottom) $a_s = a_{nn}$, $a_s = 5a_{nn}$, $a_s = 10a_{nn}$, and $a_s = 50a_{nn}$ (see legend). Also shown are hyperradial potentials $U(\rho) = \frac{\hbar^2 \ell_{\text{eff}}(\ell_{\text{eff}}+1)}{2\mu\rho^2}$ for the non–interacting limiting case (dashed) with $\ell_{\text{eff}} = 5/2$ and the unitarity limiting case (dot–dashed) with $\ell_{\text{eff}} = 1.2727$ [12, 14]. The inset plot shows the hyperradial potential for the bare neutron–neutron scattering length on a zoomed–out scale to highlight, at small $\rho$, the classically forbidden region where the wave function is evaluated.

### 3.2 4n System

It is possible to carry out a similar study of processes that would suddenly create four free neutrons in their low energy open continuum. Consider a $^4$He nucleus that is surrounded by two negative pions (or muons) in the 1$s$-orbital. If this system were created in the laboratory, probably the usual reaction would involve one pion (muon) at a time getting captured by one of the two protons in the nucleus.

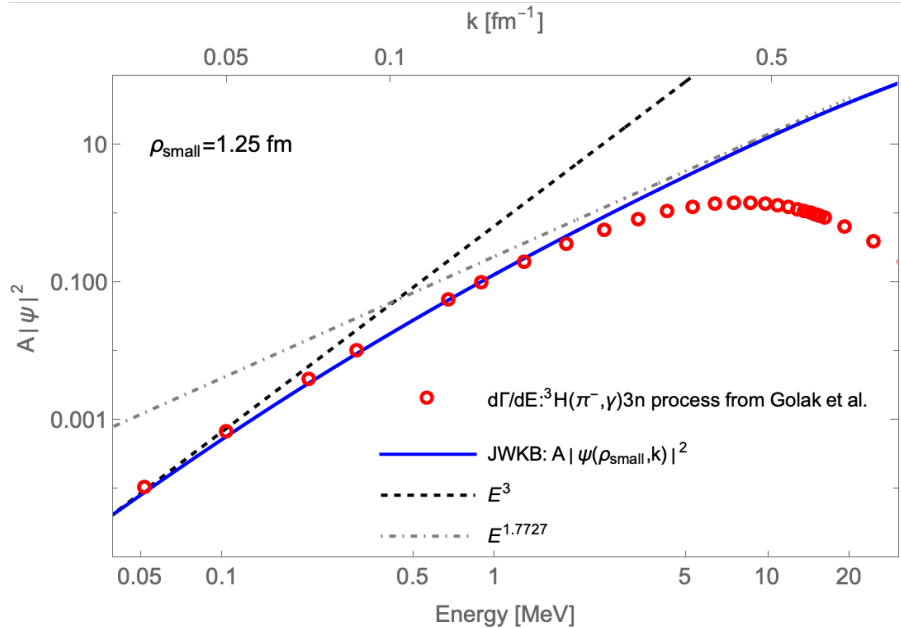

**Fig. 2** This plot shows the JWKB density versus energy, evaluated at $\rho_{\text{small}} = 1.25$ fm, for the 3n system in the $J^\pi = 3/2^-$ symmetry (dot–dashed). The JWKB analysis was performed on the lowest adiabatic hyperspherical potential curve computed in Ref. [14] The energy-dependent final state density at $\rho_{\text{small}}$ is rescaled by a constant and compared with the differential capture rate $d\Gamma/dE$ for the pion capture process $^3\text{H}(\pi^-,\gamma)3n$ [10]. At very low energies (Region *(I)*, $k \lesssim 1/|a|$), the Golak *et al.* calculation is in agreement with our hyperspherical treatment that the scaling law behavior is $E^3$, the true threshold law away from unitarity. At intermediate energies (Region *(II)*, $0.05 \lesssim k \lesssim 0.3$), the Golak *et al.* calculated rate also agrees with the hyperradial JWKB density scaling, as expected for 3 neutrons in the $3/2^-$ symmetry.

But some fraction of the time, two pions (muons) could be simultaneously captured, in the reaction $^4\text{He}(\pi^-\pi^-,\gamma\gamma)4n$ (or $^4\text{He}(\mu^-\mu^-,\nu\nu)4n$). If this process could be isolated experimentally, it would lead to the formation of 4 neutrons and the low energy threshold exponent could be studied, in analogy with the 3 neutron process that was analyzed above. We view this as a *gedanken* experiment, because its difficulty makes it unlikely to ever be actually carried out in the laboratory. But the analysis in terms of the relevant lowest energy adiabatic hyperspherical potential curve for the 4n system in the $0^+$ symmetry should give the threshold behavior and also indicate whether the unitarity-modified threshold exponent would be observed in the energy-dependent capture rate.

Figure 5 shows the lowest hyperradial potential curve for four interacting neutrons in the $J^\pi = 0^+$ symmetry, calculated using the AV8′ interaction. At large hyperradius, the potential energy curve approaches the non–interacting curve (dashed curve) given by $W_0(\rho) = \frac{\hbar^2 \ell_{\text{eff}}^{\text{NI}}(\ell_{\text{eff}}^{\text{NI}}+1)}{2\mu\rho^2}$ with $\ell_{\text{eff}}^{\text{NI}} = 5$. In the limit that the $s$–wave interaction becomes resonant, and thus the scattering length becomes infinite, $\ell_{\text{eff}}$ gets reduced to the value $\ell_{\text{eff}}^{\text{U}}$ shown in Table 1. Again, as was stressed Refs. [3, 4], this reduction can be viewed as a manifestation of Efimov physics. In the $0^+$ symmetry for the 4n system, the reduced value of the effective centrifugal constant is $\ell_{\text{eff}}^{\text{U}} = 2.017$ [12]. This limit is shown in Fig. 5 as the lower dot–dashed curve. In practice, the adiabatic hyperspherical potential curve is fitted to its asymptotic form at intermediate hyperradii, and used at large hyperradii. The $4n$ hyperradial potential curve shown in Fig. 5 has been fitted to the following form for $\rho > 10$ fm,

$$W_0^{(4n)}(\rho) = \frac{\hbar^2}{2\mu\rho^2}\left(\ell_{\text{eff}}^{\text{NI}}(\ell_{\text{eff}}^{\text{NI}}+1) + \frac{Ca_{nn}}{\rho} + \frac{C_2}{\rho^2} + \frac{C_3}{\rho^3} + \frac{C_4}{\rho^4} + \frac{C_5}{\rho^5} + \frac{C_6}{\rho^6}\right). \quad (14)$$

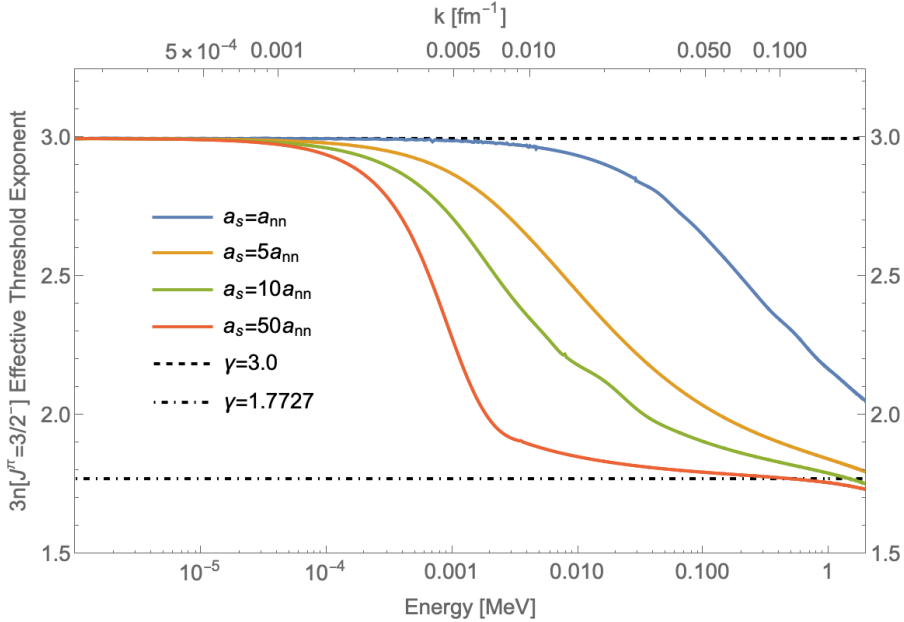

**Fig. 3** This plot shows the three–body effective threshold exponent $\gamma(k,\rho_{\text{small}})$, defined in Eq. (11) using the JWKB wave function, and evaluated at $\rho_{\text{small}} = 1.25$ fm, vs. energy for different neutron–neutron scattering lengths. The scattering length is tuned by varying the overall strength of the spin–singlet two–body interaction. The upper to lower curves are for neutron–neutron scattering lengths of $a_s = a_{nn}$, $a_s = 5a_{nn}$, $a_s = 10a_{nn}$, and $a_s = 50a_{nn}$, respectively. The horizontal lines indicate the effective threshold exponents for the non-interacting case with $\gamma = 3.0$ and at unitarity, namely $\gamma = 1.7727$. The horizontal axis at the top of the figure gives the wave number $k$. A general trend observed in these curves is that deviations of $\gamma$ from the finite interaction or "true" threshold exponent, $\ell_{\text{eff}}^{\text{NI}} + \frac{1}{2}$ that is relevant for $|a| < \infty$, become significant after $k$ reaches a value of approximately $k \gtrsim 1/|a|$.

For the case where the singlet scattering length is the bare neutron–neutron $^1S_0$ value denoted $a_{nn}$, the expansion coefficients for the $4n$ hyperradial potential are $C = 86.7(3)$, $C_2 = 66.8(4) \times 10^3$ fm$^2$, $C_3 = -1.61(2) \times 10^6$ fm$^3$, $C_4 = 2.16(4) \times 10^7$ fm$^4$, $C_5 = -1.49(4) \times 10^8$ fm$^5$, and $C_6 = 4.13(12) \times 10^8$ fm$^6$. With this hyperradial potential energy curve, the wave function in the JWKB approximation given by Eq. (8) is calculated to obtain the effective threshold exponent for the $4n$ system.

Figure 6 shows the low energy JWKB density (a), while (b) shows the effective threshold exponent $\gamma(k,\rho_{\text{small}})$ computed using the JWKB wave function (solid curve) for the $4n$ system, evaluated also at $\rho_{\text{small}} = 1.25$ fm, over an energy range from $0 < E < 3$ MeV. The two horizontal lines display the value for the threshold exponent in the two limiting cases, four non–interacting particles (dashed line) with an exponent $\gamma = 11/2$, and four resonantly interacting particles at the $s$–wave unitarity limit (dot–dashed line) with an exponent $\gamma = 2.517$. (Hammer and Son [2] cite a similar value for the unitarity exponent but with less precision, quoting values in the range 2.5-2.6.) As the energy increases, the exponent changes from the non–interacting value and decreases towards the unitarity value. Due to the nature of the AV8′ interaction, the neutron–neutron $s$–wave scattering length, while large, is not near the unitarity limit, so the threshold exponent stays well above the four–body exponent $\gamma = 2.517$ at all energies in the threshold energy regions *(I)* and *(II)*. A similar situation was seen for the three–body system in the previous section. If the neutron–neutron scattering length could be tuned to larger values on the negative side, then the effective threshold exponent would get closer to the unitarity value for smaller energies and only then would the unitarity-modified threshold exponent become observable.

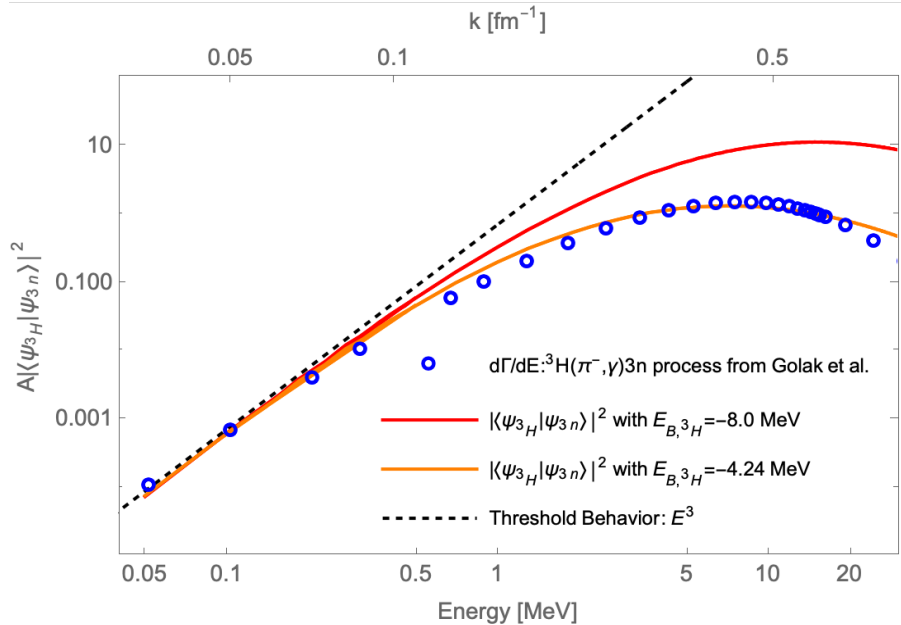

**Fig. 4** This plot shows a comparison between the capture rate $d\Gamma/dE$ for the pion capture process $^3\text{H}(\pi^-,\gamma)3n$, and the overlap of the triton wave function with the $3n$ wave function at a given energy $E$, denoted as $|\langle\psi_{3\text{H}}|\psi_{3n}\rangle|^2$. This figure demonstrates where the nodal structure of the energy-dependent final state wavefunction begins to matter, which is where Region *(III)* begins. The two solid lines indicate the overlap of the triton wave function using the Minnesota interaction with an addition of a hyperradial three–body force to modify the triton energy. The corresponding (inaccurate) triton bound–state energies are reported in the figure legend. The curvature at energies greater than 2 MeV is governed by the characteristics of the triton wave function and the triton binding energy. The lower the binding energy, the more the curve "turns over" at large energy. The dashed line indicates the true Region *(I)* threshold law behavior that scales as $E^3$. While the simplistic overlap between initial and final hyperradial wavefunctions omits the relevant transition operator for the pion capture process, the plot shows where the shape of the final state wavefunction starts controlling the energy-dependence.

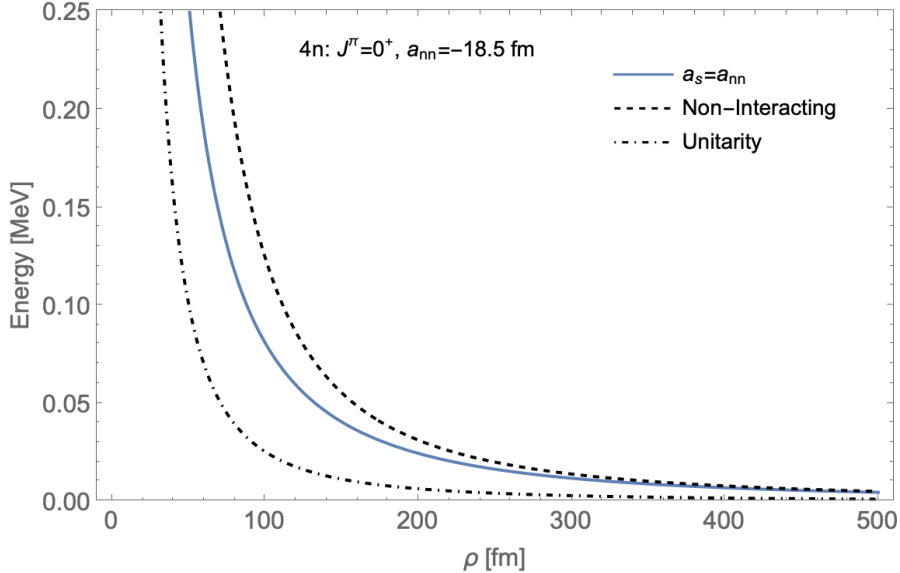

**Fig. 5** The lowest hyperradial potential curve (solid curve) for the $4n$ system in the $J^\pi = 0^+$ symmetry, computed using the AV8′ two–body interaction. The dashed line indicates the non–interacting potential energy curve with $\ell_{\text{eff}} = 5$. The dot–dashed curve indicates the $\rho^{-2}$ term in the potential energy curve at the $s$–wave unitarity limit, where there is a reduction in $\ell_{\text{eff}}$ giving $\ell_{\text{eff}} = 2.017$ (see Table 1).

## 4 Conclusion

This study addresses the near-threshold behavior close to $s$–wave unitarity for systems containing a few neutrons. The values of the modified threshold exponents expected at exact unitarity in our study are consistent with the values predicted by Hammer and Son[2] using conformal field theory, but the realistic neutron-neutron scattering length is not sufficiently close to unitarity to observe these modified threshold laws. Our analysis considers two example cases: a study of three interacting neutrons and a study of four interacting neutrons. In the study of three interacting neutrons, the JWKB wave function is calculated using the lowest three–body hyperradial potential curve in the $J^\pi = 3/2^-$ symmetry, which is the most attractive at long range. The adiabatic hyperspherical potential curves are computed using both the AV8′ nucleon–nucleon interaction and a single Gaussian representation of the interaction. The tunneling probability is then analyzed as a function of the collision energy to extract the near–threshold behavior. The tunneling probability is a measure of the probability that the three neutrons can be formed at a given energy under the angular momentum barrier following muon or pion capture by a triton or an alpha particle. The near–threshold behavior of the tunneling probability is compared directly to the threshold behavior of the differential capture rate of the muon and pion capture processes $^3\text{H}(\pi^-,\gamma)3n$ and $^3\text{H}(\mu^-,\nu_\mu)3n$. The near–threshold behavior at very low energy (Region (I)) obtained in the hyperspherical treatment agrees with the Fadeev treatment of Golak et al. and there it follows the ordinary Wigner-type power law $E^{\ell_{\text{eff}}+1/2}$ with $\ell_{\text{eff}} = 5/2$. At somewhat higher energies (Region *(II)*), the adiabatic hyperspherical potential curves at the outermost classical turning point are modified from that purely centrifugal behavior, and the effective threshold exponent $\gamma$ begins to deviate from the value $\ell^{\text{NI}}_{\text{eff}} + \frac{1}{2}$, until at higher energy still (Region *(III)*), where a short-range transition operator begins to vary from simple power law behavior as the final state wavefunction shape changes with energy. As part of our analysis, the threshold behavior near the $s$–wave unitarity limit was further analyzed by varying the $^1S_0$ $n-n$ scattering length to more negative values. From this, it is apparent that the unitarity threshold law of $E^{1.7727}$ is observed only when the scattering length is closer to negative infinity, probably 3-5 times more negative than the nominal value $a_{nn} = -18.5$ fm, as can be seen in the effective threshold exponent plotted in Fig. 3.

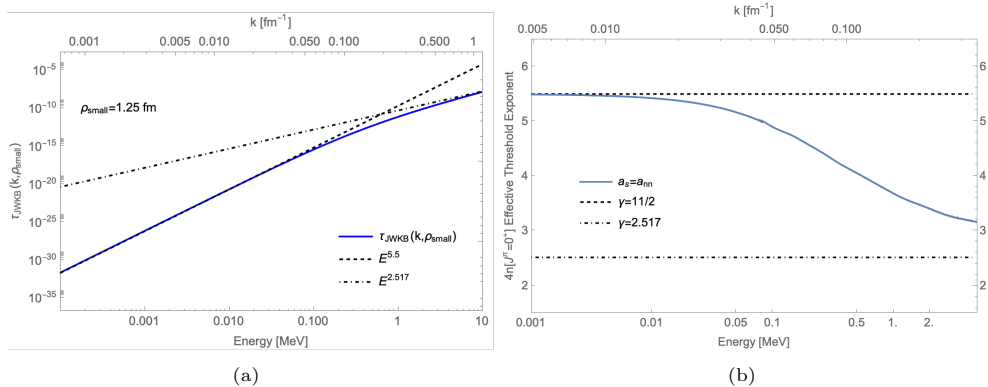

**Fig. 6** The squared JWKB tunneling integral $\tau_{\text{JWKB}}(k,\rho)$ is shown in (a) as a function of energy for the $4n$ system at a hyperradius $\rho_{\text{small}} = 1.25$ fm. The dashed and dot–dashed lines indicate the power law $E^{11/2}$ at threshold (dashed) and the unitarity power law $E^{2.517}$ (dot–dashed). The effective threshold exponent $\gamma(k, \rho_{\text{small}})$, defined in Eq. (11), for the $4n$ system in the $J^\pi = 0^+$ symmetry evaluated at $\rho_{\text{small}} = 1.25$ fm, is shown in (b). The horizontal lines present the values of the effective threshold exponent in the limiting cases of non–interacting or (finite $|a|$ at $k \ll 1/|a|$) particles (dashed) with a value of $\gamma = 11/2$ and resonantly interacting particles (dot–dashed) with a value of $\gamma = 2.517$ (see Table 1) where the $s$–wave interaction is tuned to the unitarity limit. The top axis labels the wave number $k$. Similar to the three body case, deviations from the non–interacting threshold exponent greater than $\approx 0.5$ occurs at values of the energy for which $k \gtrsim 1/|a|$.

The near–threshold behavior for the four–neutron system was also studied. The lowest adiabatic hyperspherical potential curve in the $0^+$ symmetry was used to compute the JWKB wave function and tunneling probability. The effective threshold exponent was then calculated and compared to the non–interacting or finite $|a|$ value of $\gamma = 11/2$ and the value at $s$–wave unitarity, $\gamma = 2.517$. Our results show the threshold exponent decreasing in Region *(II)* from the non–interacting limit value in Region *(I)*; however, $\gamma$ does not appear to reach the unitarity value, deviating by around 20% at the upper end of Region *(II)*. The four-neutron results show behavior analogous to our three-neutron results, and demonstrate that while the true neutron–neutron scattering length is large and negative, it leads to physics (i.e. transition amplitude threshold exponents) that still differ significantly from unitarity in the near–threshold limit.

Only when the scattering length is tuned to much larger values does the near–threshold exponent attain its unitarity value over an appreciable energy range.

**Acknowledgements.** This work is supported in part by the National Science Foundation, Grant No. PHY-2207977. This work was additionally partly supported by the National Science Centre, Poland, under Grant No. IMPRESS-U 2024/06/Y/ST2/00135 and by the Excellence InitiativeResearch University Program at the Jagiellonian University in Krakhe numerical calculations were partly performed on the supercomputers of the JSC, Jülich, Germany.